\documentclass[10pt, conference, compsocconf]{IEEEtran}
\usepackage{ifpdf}
\usepackage{cite}
\ifCLASSINFOpdf
  \usepackage[pdftex]{graphicx}
  \graphicspath{{../pdf/}{../jpeg/}}
  \DeclareGraphicsExtensions{.pdf,.jpeg,.png}
\else
  \usepackage[dvips]{graphicx}
  \graphicspath{{../eps/}}
  \DeclareGraphicsExtensions{.eps}
\fi
\usepackage[cmex10]{amsmath}
\usepackage{algorithmic}

\usepackage{array}

\usepackage{mdwmath}
\usepackage{mdwtab}
\usepackage{eqparbox}

\usepackage[tight,footnotesize]{subfigure}
\usepackage[caption=false,font=footnotesize]{subfig}

\usepackage{stfloats}
\usepackage{multirow}
\usepackage{bigstrut}
\usepackage{hhline}
\usepackage{url}
\makeatletter
\def\@cite#1#2{[\if@tempswa #2 , \fi#1]}
\makeatother

\begin{document}
\title{Speculative Symbolic Execution}


\author{\IEEEauthorblockN{Yufeng Zhang, Zhenbang Chen, Ji Wang}
\IEEEauthorblockA{National Laboratory for Parallel and Distributed Processing\\
Department of Computing Science, National University of Defense Technology\\
Changsha, China\\
Email: \{yufengzhang, zbchen\}@nudt.edu.cn, jiwang@ios.ac.cn}
}


%Telephone: (800) 555--1212, Fax: (888) 555--1212}


\maketitle

\begin{abstract}
Symbolic execution is an effective path oriented and constraint based program analysis technique. Recently, there is a significant development in the research and application of symbolic execution. However, symbolic execution still suffers from the scalability problem in practice, especially when applied to large-scale or very complex programs. In this paper, we propose a new fashion of symbolic execution, named \emph{Speculative Symbolic Execution} (SSE), to speed up symbolic execution by reducing the invocation times of constraint solver. In SSE, when encountering a branch statement, the search procedure may speculatively explore the branch without regard to the feasibility. Constraint solver is invoked only when the speculated branches are accumulated to a specified number. In addition, we present a key optimization technique that enhances SSE greatly. We have implemented SSE and the optimization technique on Symbolic Pathfinder (SPF). Experimental results on six programs show that, our method can reduce the invocation times of constraint solver by $21\%$ to $49\%$ (with an average of $30\%$), and save the search time from $23.6\%$ to $43.6\%$ (with an average of $30\%$).
\end{abstract}


% \fi
\IEEEpeerreviewmaketitle

\section{Introduction}
Symbolic execution (SE) is a basic program analysis technique that was proposed more than thirty years ago \cite{king1976symbolic}. Recently, SE draws renewed interests both from academia and industry partly due to the impressive progress in constraint solving, related algorithms and computation power \cite{cadar2011ICSEsymbolic}\cite{puasuareanu2009newtrends}\cite{puasuareanu2007HVCsymbolic}. Instead of executing programs with concrete inputs, symbolic execution feeds programs with symbolic ones, meaning that a symbolic input could initially take any value of the specific type. Assignment statements are interpreted as the manipulations of symbolic expressions. When encountering a branch statement, the process forks and both of the branches are taken. On each path, the process maintains a set of constraints called \emph{path condition} which must hold along that path. For each branch, the path condition is updated according to the corresponding condition and submitted to a constraint solver to check the satisfiability. In the context of test generation, when a path ends or a bug is found, the path condition can be solved to get a test case. For deterministic programs, the same execution path or the same bug can be replayed by feeding such test case as input.
Basically, symbolic execution attempts to achieve automatic code comprehension by walking through the path space of a program. Providing that all the path conditions can be solved successfully, symbolic execution could cover all the behaviors of the program.

In the past years, symbolic execution has shown a great promise in the application to automated test generation, proving program properties, bug detection and so on \cite{puasuareanu2009newtrends}. However, in practice, the scalability problem is still one of the main obstacles in applying symbolic execution to large-scale programs. This issue mainly stems from two closely related reasons: path explosion phenomenon and constraint solving overhead. There exists an exponential relationship between the number of conditions and the paths of the program, making exploring the whole path space infeasible for large-scale programs. Constraint solving is the most dominant in the running time of SE. When exploring deep paths, the path condition may be very complex, and even unsolvable. In addition, constraint solving overhead is almost always aggravated by the path explosion phenomenon.

To alleviate the constraint solving overhead of SE, many techniques have been proposed. In many symbolic execution systems, query optimization techniques are employed to reduce the complexity of queries and query times. For example, \emph{counterexample caching} stores unsatisfiable path conditions as counterexamples to reuse previous solving results \cite{cadar2008klee}. \emph{Constraint independence} splits a constraint set into independent ones, aiming to get the related constraint set and increase the cache hit rate \cite{cadar2008klee}\cite{cadar2008exe}\cite{sen2005cute}\cite{godefroid2008sage}. \emph{Concretization} reduces complex constraints (such as nonlinear constraints \cite{puasuareanu2011mixedsolving}) into simpler ones, and is heavily used in concolic execution \cite{sen2005cute}\cite{godefroid2008sage}\cite{godefroid2005dart}.

Although these effective techniques improve the performance of symbolic execution greatly in practice, constraint solving is still the most dominant in symbolic execution. According to the experiments of KLEE \cite{cadar2008klee}, $40\%\sim 90\%$ of the whole running time is spent on constraint solving. In the experiments of Cloud9 \cite{ciortea2010cloud9}, constraint solving consumes more than half of the total execution time. In some experiments in S$^2$E \cite{chipounov2011s2e}, almost all the running time is dominated by the constraint solving.

%%zyf:no more data for solving overhead now

This paper proposes a new fashion of symbolic execution, named \emph{Speculative Symbolic Execution} (SSE), which speeds up symbolic execution by reducing the invocation times of constraint solver, and hence improves the scalability of symbolic execution. Unlike pure symbolic execution, which invokes the constraint solver immediately when a path condition is updated, in SSE, when a branch instruction is encountered, the path condition is updated accordingly, but the constraint solver is not necessarily invoked. The search procedure may advance along the path without the determination of feasibility until the unsolved path conditions are accumulated to a specified number. If the current visiting path is feasible, the procedure continues; otherwise, it backtracks.

Intuitively, SSE takes branches optimistically as feasible ones. Path conditions are submitted to constraint solver in batches, not one by one as in pure symbolic execution. When speculation succeeds, multiple invocations of constraint solvers are replaced by one invocation. When speculation fails, a backtracking mechanism will find the first bad branch that makes the speculation fail. Basically, the more feasible branches in the path space, the better SSE performs.

%In this paper, we research the combination of speculation with depth first search (DFS) algorithm.
In this paper, we give out the details of SSE algorithm and discuss its effectiveness. We also propose an optimization technique, named \emph{Absurdity Based Optimization}, which is simple but very effective in practice. For programs with a high ratio of infeasible branches in the path space, this optimization can reduce the times of invoking constraint solver significantly. To some extent, our optimization is complementary to SSE, and can also be applied to pure symbolic execution.

The contribution of this paper is three-fold.

Firstly, we propose speculative symbolic execution, a new fashion of symbolic execution, to extend the scalability of classical symbolic execution by reducing the invocation times of constraint solver. We also propose absurdity based optimization technique to improve the reduction further.

Secondly, we have implemented SSE and the optimization on top of Symbolic Pathfinder \cite{pasareanu2008combining} to extend the scalability of this symbolic execution system.
%PathFinder extension \cite{pasareanu2008combining} of JPF.

Finally, to evaluate the effectiveness of our method, we have conducted several experiments and find a new characteristic of the path spaces of programs. The experimental results show that our approach can save the search time from $23.9\%$ to $43.6\%$ (with an average of $30\%$). Based on these results, we also investigate how to make our approach work best when applied to real world programs.

The remainder of this paper is organized as follows. Section 2 introduces the background and shows the basic idea of SSE by motivating examples. Section 3 elaborates the algorithm of SSE and the absurdity based optimization technique. Section 4 presents our implementation on SPF and reports the experimental results. Finally, Sections 5 and 6 discuss the related work and conclude.

\section{Overview}

In this section, we describe how SSE works and why SSE is better than pure SE by motivating examples.

\subsection{Background: Symbolic Execution}
Essentially, symbolic execution feeds programs with symbolic values as inputs and outputs the result as functions of symbolic values. A search procedure is employed to systematically traverse the path space of a program by maintaining symbolic program states. A symbolic state includes the symbolic values of program variables, a path condition and a program counter \cite{king1976symbolic}. The path condition is a boolean formula that contains the constraints which the inputs should satisfy if they drive the program along the current path. Operations of variables are interpreted as the manipulations of symbolic expressions. When encountering a branch instruction, both of the branches are taken. For each branch, the corresponding condition is added into the path condition and a constraint solver is invoked to check the satisfiability of the new path condition. The process advances along feasible branches until the path ending is reached. Finally, the generated symbolic states form a \emph{symbolic execution tree}.
\begin{figure}[b]
\centering
\includegraphics[width=1.2in]{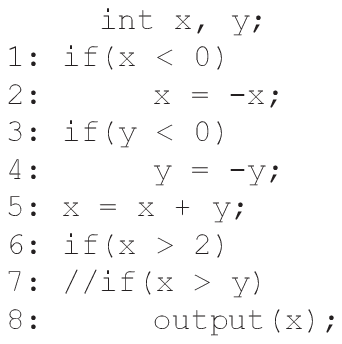}
\caption{An Example Program and Its Execution Tree}
\label{Fig:Fig1}
\end{figure}

\begin{figure*}[t]
\centering
\includegraphics[width=5in]{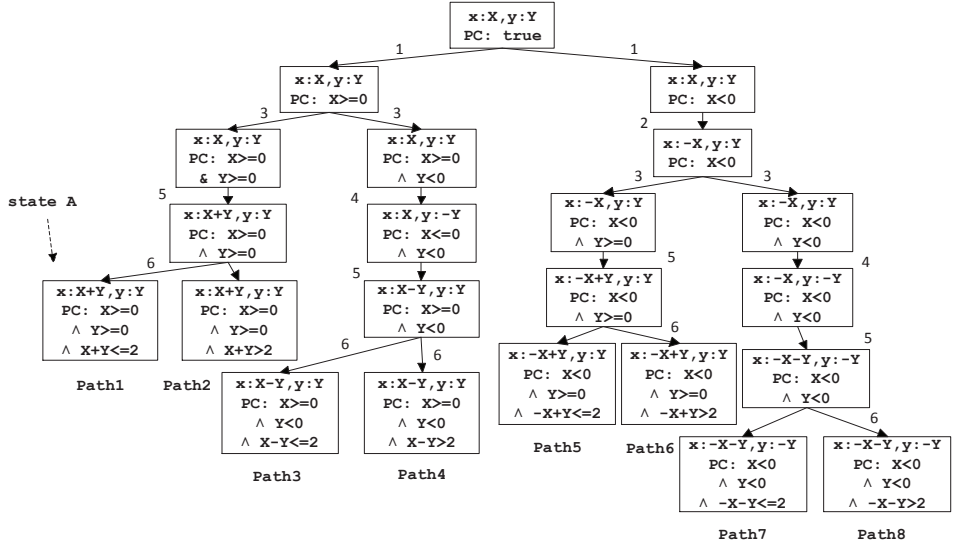}
\caption{An Example Program and Its Execution Tree}
\vspace{-0.2cm}
\label{Fig:Fig2}
\end{figure*}

Take the program in Figure \ref{Fig:Fig1} for example. It computes the sum of the absolute values of two integers and outputs the result if the sum is greater than $2$. Initially, the inputs are represented as two symbols: $X$ and $Y$, and the path condition is $\langle true\rangle$. Execution path forks when meeting the branch statement \verb|if(x<0)|. The constraints $\langle X\geq0\rangle$ and $\langle X<0\rangle$ are added to the path conditions of the two paths respectively. A constraint solver is invoked to check the feasibility of these two paths, both of which here are feasible. Figure \ref{Fig:Fig2} shows the final execution tree, in which symbolic states are represented as nodes.
%In all, there are $8$ paths in the execution tree. At the end of each feasible path, the constraint solver can be invoked to generate test cases, which is endowed to drive the program along the corresponding path if fed as input.

In this paper, we focus on how the constraint solver is invoked during symbolic execution. We choose the commonly used \emph{depth first search} (DFS) in our illustration.

\begin{figure}[b]
\centering
\includegraphics[width=3in]{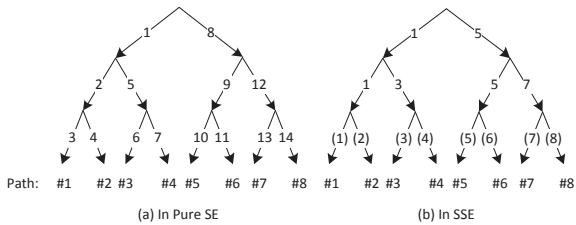}
\caption{Constraint Solving in DFS}
\label{Fig:Fig3}
\end{figure}

Figure \ref{Fig:Fig3}(a) shows the path space of the example program with the same layout in Figure \ref{Fig:Fig2}. The left side of a node corresponds to the \emph{false} side of the branch statement. The number $n$ marked on a branch means that the feasibility of the branch is determined in the \emph{n}-th invocation of the solver. Totally, $14$ times of constraint solving are needed.

\subsection{Motivating Examples of SSE}
When encountering a branch statement, SSE may advance along the two branches without checking the feasibility.
%Sequentially, the path condition of the current visiting symbolic state may contain several unchecked constraints.
The constraint solver is invoked only when the number of unchecked branches reaches a specific number, say \emph{max speculation depth}. If the constraint solver gives a positive result, it means that the speculation succeeds. Otherwise, we need backtrack to the last feasible branch. Now we present how speculation reduces solving times in a DFS manner with the example in Figure \ref{Fig:Fig1}.

The initial symbolic state of the program under SSE is the same as that under SE. Assuming that the max speculative depth is set as $3$, for branch statement \verb|if(x<0)|, the procedure advances along the \verb|else| side without checking feasibility. Branches of the statement \verb|if(y<0)| are handled similarly. When the procedure takes the \verb|else| branch of statement \verb|if(x>2)| speculatively, the max speculation depth is reached, therefore a constraint solver is invoked. Since the path segment from root to \emph{state A} (in Figure \ref{Fig:Fig2}) is executed speculatively, we call this segment a \emph{speculation segment}. As a result, only one time of constraint solving is enough to know the feasibility of the three branches on path \#1. As shown in Figure \ref{Fig:Fig3}(b),
 %illustrates when and how the constraint solver is invoked in speculative DFS.
 %the bracketed number shows where the constraint solver is invoked.
the number $n$ associated on a branch demonstrates the feasibility of the branch is known in the \emph{n}-th solving. The invocations of constraint solver only occur at the branches marked with bracket numbers. In all, only $8$ queries are needed, saving nearly half of that in pure SE.

\begin{figure}[t]
\centering
\includegraphics[width=2.8in]{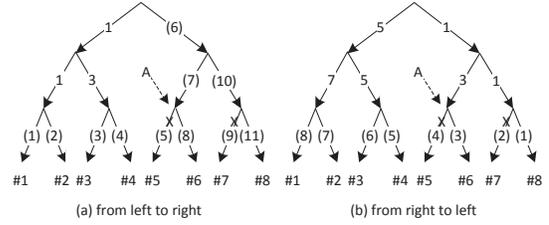}
\caption{Constraint Solving in SSE by DFS With Backtracking}
\label{Fig:Fig4}
\vspace{-0.2cm}
\end{figure}

Now consider commenting line 6 and uncommenting line 7 in the example program in Figure \ref{Fig:Fig1}. Path \#5 and \#7 would be infeasible. In Figure \ref{Fig:Fig4}, they are marked with a cross. In this case, the number of constraint solving under pure SE is still $14$. In SSE, as shown in Figure \ref{Fig:Fig4}(a), the result of the $5$th time of solving with path condition $\langle X>0 \wedge Y\leq 0 \wedge X \leq Y\rangle$ is \emph{unsat}. In the sequel, the backtracking mechanism analyzes the current speculation segment (\emph{i.e.}, from root to \emph{point A}) in a binary search way to find the first infeasible branch, which spends two extra times (6th and 7th) of constraint solving. Then the procedure backtracks to \emph{point A} and continues on path \#6. Constraint solving on path \#7 is similar to that on path \#5. Finally, $11$ times of constraint solving with $9$ \emph{sat} and $2$ \emph{unsat} are performed, saving $3$ out of $14$ in pure SE.

It is worth noting that the result of SSE is related to the order in which the path space is explored. Consider exploring the path space from right to left, \emph{i.e.}, exporing the true side of a branch statement first. As shown in Figure \ref{Fig:Fig4}(b), only $8$ times of constraint solving is enough.

\section{Speculative symbolic execution}
%SSE attempts to reduce the times of constraint solving when exploring the program path space.
One can imagine using different search styles in SSE. In this section, we present the \emph{speculative DFS} algorithm that combines speculation and DFS, and the absurdity based optimization. Then we discuss the effectiveness of our approach.

\subsection{Speculative DFS Algorithm}
\begin{figure}[t]
\centering
\includegraphics[width=3in]{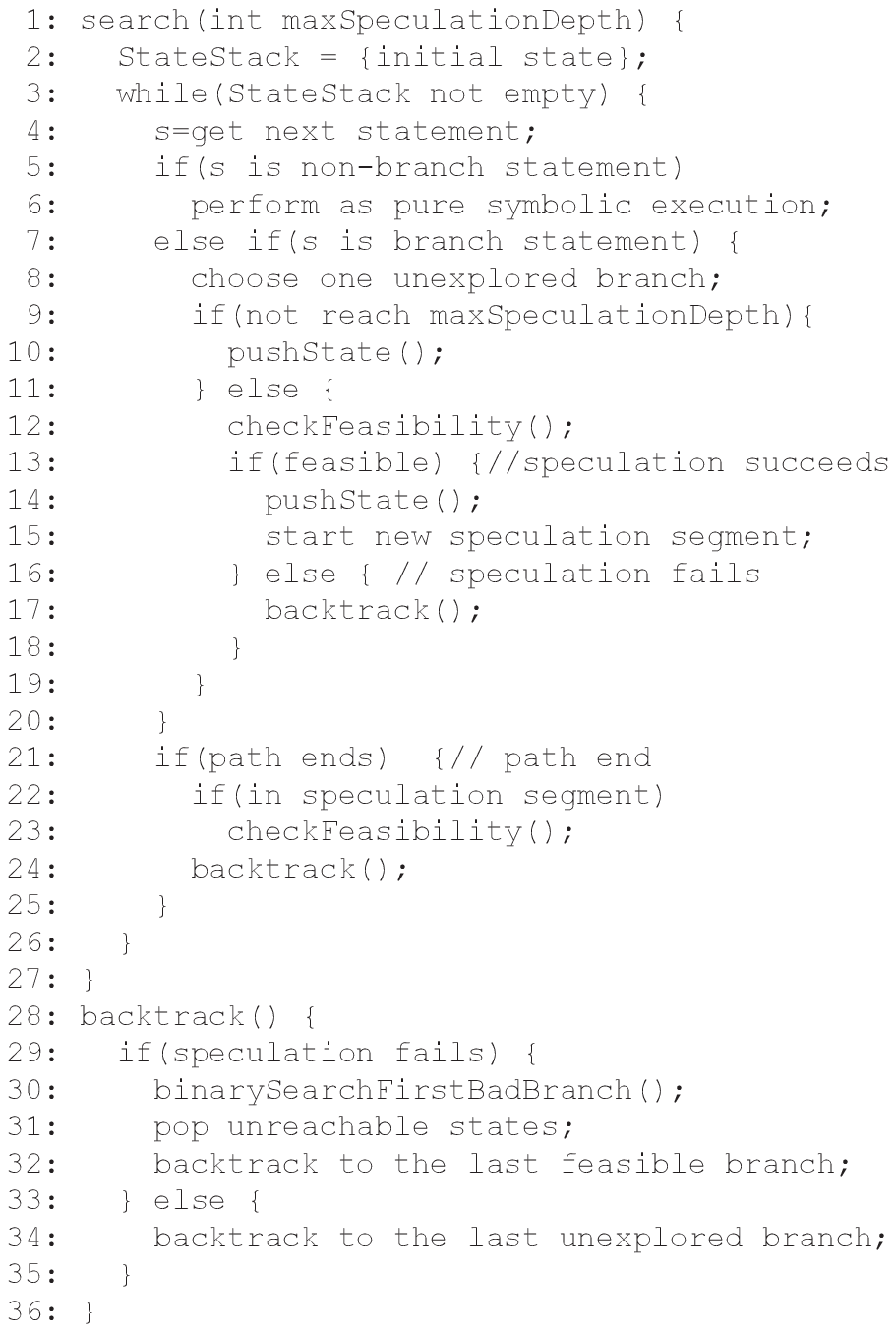}
\caption{Speculative DFS Algorithm}
\vspace{-0.6cm}
\label{Fig:Fig5}
\end{figure}

Figure \ref{Fig:Fig5} shows the algorithm of speculative DFS, including the main \verb|search| procedure and the \verb|backtrack| procedure. The algorithm traverses the path space of program by DFS and performs speculation with a specially designated backtracking mechanism. A \verb|StateStack| is maintained to store the symbolic states on the current path. Initially, the initial symbolic state of the program is pushed into the \verb|StateStack|. The \verb|while| loop expands the top element of the \verb|StateStack| until the stack is empty. The procedure forwards by symbolically executing the next statement of the top state in the \verb|StateStack| repeatedly. For a non-branch statement, our algorithm performs identically with pure SE. When processing a branch statement, if the current speculation depth has not reached the \verb|maxSpeculationDepth|, the branch is taken without checking feasibility and a new state with updated path condition is pushed into the \verb|StateStack| directly as shown in line 10. Otherwise, as shown in line 12, function \verb|checkFeasibility()| checks the satisfiability of the current path condition. If the result is \emph{sat}, the current state is pushed into the \verb|StateStack| and a new speculation segment starts. If the result is \emph{unsat}, the \verb|backtrack()| procedure cuts the infeasible branches away. The procedure backtracks when reaching the end of a path. According to the feasibility of the last speculation segment on the path, the \verb|backtrack()| procedure performs differently. Note that, when the \verb|maxSpeculationDepth| is set as $1$, this algorithm is equivalent to pure SE.

%In pure SE, when the process reaches the end of a path, it backtracks to the last unexplored branch, if any. While i
%Speculative DFS may execute the last statement of a path speculatively. Therefore, as shown in line 23, function \verb|checkFeasibility()| is invoked. According to the result, the algorithm performs different backtracking.

The backtracking procedure performs differently in different cases to suit for the context of speculation. For a failed speculation with $k$ branches, the speculation segment before the last branch (already known as infeasible) is analyzed to find the first infeasible branch. We adopt the binary search strategy for its stable performance in different cases,
%Assuming that a speculation segment is formed by $k$ branches $b_1, ..., b_k$, the corresponding path conditions are $p_1, ..., p_k$, where $p_{i+1}=p_i\wedge c_{i+1}\ (i<k\ \text{and}\ c_{i+1} \ \text{is from}\ b_{i+1})$. In line 30, we perform binary search between $p_1$ and $p_{k-1}$ to find the first bad path condition.
which needs at most $\lceil \log_2(k-1) \rceil$ times of constraint solving. Line 34 deals with another case when the path ends with a reachable state, the procedure backtracks to the last unexplored branch.

%We define the correctness of SSE as
%
%``\emph{Speculative symbolic execution generates the same execution tree as pure symbolic execution in the end}''.
%
%For the sake of space, the proof of the correctness of speculative DFS algorithm is available in the accompanying long version of this paper \cite{yufengzhang2012sse}.

\subsection{Eliminating False Alarms}
\begin{figure}[h]
\centering
\includegraphics[width=1.7in]{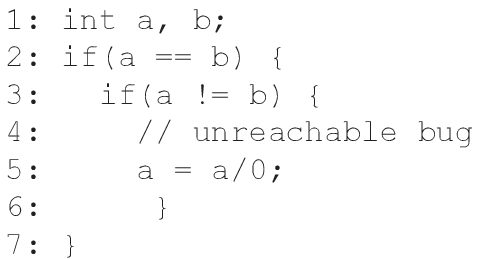}
\caption{Another Example Program}
\label{Fig:Fig6}
\vspace{-0.2cm}
\end{figure}

Although SSE generates the same execution tree as pure SE, in practice, bugs located in dead code may cause SSE yielding different analysis results from pure SE. Consider the example shown in Figure \ref{Fig:Fig6}, line 5 contains a `divide-by-zero' bug; however, it is unreachable since its path condition $\langle a=b\wedge a\neq b\rangle$  is unsatisfiable. In SSE, providing that the two branch statements in line 2 and line 3 are in a same speculation segment, line 5 would be executed without a prior determination of its reachability. Hence a false alarm will be reported.

Technically, this issue can be simply addressed via checking the reachability of the potential bug point just before generating the bug report. It is necessary to point out that the exceptions caused by constraint solving (such as caused by the constraints beyond the ability of constraint solver) should be handled carefully, because a repeated reachability checking would trigger the same exception again. In this situation, the speculation segment should be checked carefully to find the first solvable and feasible branch, if any.

%If the bug point is reachable, generate a bug report and otherwise backtrack.

%Commonly, symbolic execution is applied in test generation, runtime error detection, program property checking and so on.
%The problem mentioned above should be handled differently when symbolic execution is used under different contexts. For example, reachability checking before reporting a property violation is necessary when symbolic execution is used to check safety of programs, but does not exist when used for test generation.

%if symbolic execution is used to check the safety of the program, a property violation may be located in the dead code, so reachability checking before reporting property violation such error report is necessary. While in the context of test generation, there is no need to do this.

%It is worth noting that, exceptions caused by constraint solving should be handled carefully. Consider that an exception is caused by constraint solving at the end of a speculation segment because the constraint cannot be handled by adopted constraint solver. In such case, the speculation segment should be checked carefully to find the first solvable and satisfiable branch, if any.

\subsection{Correctness}
We define the correctness of SSE as

``\emph{Speculative symbolic execution generates the same execution tree as pure symbolic execution in the end}''.

Here we only give an informal description of the correctness of the speculative DFS algorithm.

With respect to the states of the generated execution tree, speculative DFS only differs from DFS under pure SE on that it may touch states that are unreachable in pure SE. Therefore, to show the correctness of speculative DFS, it suffices to prove the following two points: all the states with unsatisfiable path condition touched by speculative DFS will be finally cut away from the execution tree and all the cut states have unsatisfiable path condition.

On one hand, suppose that state $s$ is an unreachable state in pure SE but touched in speculative DFS. There exists a time that state $s$ is the current visited state (\emph{i.e.}, at the top of \verb|StateStack|). Let $s_1, ..., s_k (k \geq 1 \wedge s_k = s) $ be the corresponding speculation segment ending with state $s$. There are the following four cases need to consider in the \verb|while| loop.
\begin{itemize}
	\item \emph{\textbf{Case 1: State s is the end of a path}}. Line 23 in Figure \ref{Fig:Fig5} checks the feasibility of the current state and gets a negative result. Then in the backtracking procedure, line 30 analyzes the speculation segment and line 31 and 32 backtracks to the last feasible branch. All the states in $s_1,...,s_k $ with unsatisfiable path condition would be cut off from the execution tree.
	\item \emph{\textbf{Case 2: The next statement of state s is not a branch statement}} and,
	\item \emph{\textbf{Case 3: The next statement of state s is a branch statement and the max speculation depth has not been reached}}. Speculation segment  $s_1,...,s_k$  will be expanded by the \verb|while| loop in Figure \ref{Fig:Fig5} without determining the feasibility until a path end or the max speculation depth is reached. Suppose the expanded speculation segment is $s_1,...,s_k,...,s_m$. If $s_m$ is a path end, the argument is similar as case 1. Otherwise, $s_m$ reaches the max speculation depth. Line 12 checks the path condition of $s_m$. Since $s_k$ is unreachable, $s_m$ must be also unreachable. Then line 17 invokes the backtracking procedure and sequentially line 30 finds the unreachable states in $s_1,...,s_k,...,s_m$, which are in turn cut off in line 31.
    %\item \emph{\textbf{Case 4: The next statement of state s is a branch statement and the max speculation depth is reached}}. The argument is similar as that in case 3.
\end{itemize}

Thereby, we claim that all the states with unsatisfiable path condition touched by speculative DFS will be finally cut away from the execution tree.

On the other hand, the only place in our algorithm where states are cut off from the execution tree is line 31. Before that, line 30 has distinguished the reachable states from the unreachable ones, so we claim that all the cut states have unsatisfiable path condition.

\subsection{Feasibility}
The only possible thing that brings risk to the feasibility of SSE is that SSE executes dead code which are never executed in pure SE.
Our backtracking mechanism guarantees that all the infeasible states will be cut away from the execution tree. However, in practice, there may exist some speculatively executed dead codes that bring influence to the unbacktrackable components of the system (such as updating a database). In such case, when the program behaviors are impacted by these components, SSE may get a different result from pure SE. In fact, for such kind of programs, pure SE may not work either.
%One typical kind of such code is those which bring influence to the unbacktrackable components in the system, such as changing a database or accessing the network and so on. If the program behaviors depend on the status of these components, SSE would perform differently as pure SE.
%When such code is guarded by unsatisfiable path condition but executed speculatively in SSE, the system would behave differently as we want.
 %One typical kind of such code is those could crash down the symbolic executor. Consider a piece of memory manipulation code that triggers a severe computer system hole and makes the computer system crash down.
%A program containing such code as dead code can be analyzed by pure symbolic execution successfully, but may be not by our approach. Another case is that the speculatively executed dead code brings influence to the environment, such as changing a database or accessing the network and so on. If the program behaviors depend on the status of the environment, our approach may get a wrong result. In fact, for such kind of programs, pure symbolic execution may doesn't work either.

%In summary, dead code that introduces unbacktrackable influence to the system may bring risks to SSE.
This issue can be addressed by blocking the influence of speculatively executed instructions. One typical technique is providing appropriate support for symbolic execution (such as environment modeling \cite{cadar2008klee}) to make the system more backtrackable.

\subsection{Absurdity Based Optimization}
SSE treats an unexplored branches as feasible one at its first glance and backtracks when a speculation fails. This feature implies that the more feasible branches in the execution tree, the better SSE performs. Meanwhile, this feature also implies that SSE is not good at handling the programs with a high ratio of infeasible branches since too many backtrackings might negate the benefits brought by successful speculation. To address this problem, we propose a simple but effective optimization, \emph{absurdity based optimization}, which is complementary to SSE for its effectiveness on the programs with a high ratio of infeasible branches. This optimization is based on the following proposition.

\newtheorem{SSEtheorem}{Proposition}
\begin{SSEtheorem}
Regardless of runtime errors, given a reachable branch statement, at least one of its branches is feasible.
\end{SSEtheorem}

This proposition comes from the well-known \emph{Reductio AD Absurdum} in first order logic \cite{enderton1972mathematical}, which says that if  $\Gamma;\varphi$ is inconsistent, then $\Gamma \models\neg\varphi$, where  $\Gamma$ is a set of well-formed formulae (wff) and $\varphi $  is a wff. In the context of symbolic execution, for instance, let state $s$ be a reachable state with a satisfiable path condition $\langle \varphi_1 \wedge ... \wedge \varphi_n\rangle$. Suppose the next statement is a two-choice branch statement, say \verb|if(|$\phi$\verb|)|, where $\phi$ is a boolean condition. If the search procedure has explored the \verb|then| branch and find that it is infeasible, \emph{i.e.}, the constraints set ${\{\varphi_1,...,\varphi_n}\}$ and $\phi$ is inconsistent, then we can deduce that $\varphi_1,...,\varphi_n \models\neg \phi$. Therefore, without querying the constraint solver, we know that the \verb|else| branch is feasible.

\begin{figure}[t]
\centering
\includegraphics[width=1.7in]{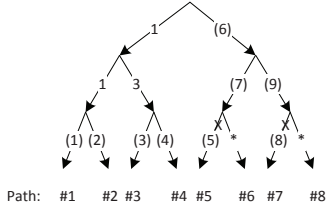}
\caption{Speculative DFS With Absurdity Based Optimization}
\vspace{-0.3cm}
\label{Fig:Fig7}
\end{figure}

This simple optimization is applicable both to pure and speculative symbolic execution. In practice, most of the branch instructions used in programs only have two choices. Therefore, as soon as an infeasible branch is explored before its counterpart, one invocation time of constraint solver can be saved. A high ratio of infeasible branches in the path space can provide many chances to perform this optimization. Consider the example in Figure \ref{Fig:Fig1} (comment line 6 and uncomment line 7), if applied with our optimization, the 8th and 11th times of constraint solving are unnecessary. As shown in Figure \ref{Fig:Fig7}, branches where constraint solving is saved are marked with asterisk.

Absurdity based optimization is also related to the order in which the execution tree is explored. If the path space in Figure \ref{Fig:Fig7} is explored from right to left, since no infeasible branch is explored before its counterpart, no information can be used to perform optimization. Thereby, we always attempt to explore the infeasible side first in practice.

\subsection{Discussion}
In this subsection, we first explain the benefits and cost of SSE, then we discuss what factors influence the effectiveness of SSE, and finally, we take a theoretical analysis on the speculative DFS algorithm.

The benefit brought by SSE is the saved constraint solvings when speculation succeeds. A successful speculation on a speculation segment of length $k$ only need once constraint solving, saving $k-1$ times compared with pure SE.

The cost of our approach lies in failed speculations. Consider a speculation segment with  $k$ branches $b_1,...,b_i,...,b_k(1\leq i\leq k)$, where branches after  $b_i$ (including $b_i$) are infeasible ones, the corresponding path conditions are $p_1,...,p_i,...,p_k$.
%In pure SE, the instructions before $b_i$ are executed and constraint solving is performed for $p_1 \sim p_{i-1}$.
In SSE, the instructions between  $b_i$ and $b_k$ are executed speculatively, which consumes extra time and memories. In addition to the first time of solving on $p_k$, binary search between $p_1 \sim p_{k-1}$ to find backtracking point needs at most $\lceil \log_2 (k-1)\rceil$  times of queries. This may be more expensive than solving for path conditions $p_1 \sim p_i$ in pure SE when $i$ is small.

The effectiveness of our approach is influenced by the characteristics of the program under analysis.
Specially, there are the following factors:
\begin{itemize}
	\item \emph{The ratio of infeasible branches in the path space}. SSE is suitable for the programs with a high ratio of feasible branches. For the programs with a high ratio of infeasible branches, SSE can be improved by the absurdity based optimization. Generally, this factor is the most important one.
	\item \emph{The shape of the path space}. SSE is also related to the shape of the path space. For example, the continuous branches on the same direction (\emph{i.e.}, all left turning or right turning) in the execution tree could increase the success rate of speculation.
	\item \emph{The exploration order over the path space}. As discussed before, both SSE and optimization depend on the exploration order.
    \item \emph{The complexity of path conditions}. The reduction of constraint solving for complex constraints can make SSE more useful.
	\item \emph{The proportion of the constraint solving time in the total running time of SE}. We only attack the constraint solving part of SE. Therefore, the proportion of constraint solving time in the total running time of SE influences our ultimate goal.
\end{itemize}

%Because of the irregularity of the program path spaces, it is hard to take a rigorous theoretical evaluation of speculative SE on an arbitrary execution tree.
The upper bound of speculative DFS algorithm is specified by the following proposition.

\begin{SSEtheorem}
The times of constraint solving in speculative depth first search are larger than half of that in pure symbolic execution.
\end{SSEtheorem}

The proof of Proposition 2 is shown in the appendix. Specially, when a path space is a full binary tree with height $n$ (the number of branches in the longest path), in pure SE, the times of constraint solving is $2^{n+1}-2$ (equal to the number of branches in the tree). While in speculative DFS, let $k$  be the max speculation depth, then the times of constraint solving $T^k_n$ can be quantified by the following equation:
\begin{displaymath}\tag{1}
T^k_n=
\begin{cases}
2^n &  (n \leq k)\\
2^n + \dfrac{2^n-2^{(n\% k)}}{2^k-1} & (n > k)
\end{cases}
\end{displaymath}

The proof of Equation (1) is shown in the appendix. For a full binary tree, speculative DFS performs best when $n \leq k$, saving nearly a half of the constraint solving times. When $n>k$, our algorithm gets better with the increase of $k$.

Speculative DFS performs worst when the execution tree only consists of a single path. In this case, although too many backtrackings affect the performance, our optimization technique can help to improve SSE. It is hard to take a precise analysis for the worst case because of the irregularity of the path spaces of programs.

In fact, both of the best case and worst case hardly happen in practice, more experimental evaluation is described in the next section.

\section{Implementation and Experimental Evaluation}

\subsection{Implementation}
We have implemented the speculative DFS algorithm and the absurdity based optimization on top of Symbolic PathFinder (SPF) \cite{pasareanu2008combining} with Java PathFinder (JPF) v6.0 \cite{JPF}\cite{visser2003model}. JPF is an open source model checker for Java bytecode. It mainly consists of a Java Virtual Machine to support state storing, state matching and backtracking, as well as an adaptive search engine to systematically explore program states. Symbolic PathFinder (SPF) is built as an extension of JPF. SPF implements symbolic version semantics for Java bytecode instructions and uses JPF to systematically explore the execution tree of program under analysis. The features of our implementation are as follows.
\begin{itemize}
	\item \textbf{New search strategy}. We have implemented the speculative DFS algorithm as a new search strategy, named \verb|SpeculativeSegmentDFSearch|, to explore the execution tree of a program speculatively. The backtracking mechanism in Figure \ref{Fig:Fig5} is employed in the new search strategy.
%A global variable \verb|curSpecuDepth| is maintained to indicate the speculation depth of the current state.
	\item \textbf{New choice generator}. We have designed a new class \verb|SpecuPCChoiceGenerator|, which is inherited from \verb|PCChoiceGenerator|. The new choice generator is utilized to help backtracking and performing the absurdity based optimization.
	\item \textbf{New semantics of branch instructions}. To Support speculative execution, the semantics of branch instructions are adapted. Each branch instruction generates an instance of class \verb|SpecuPCChoiceGenerator|. Speculation is performed according to the current speculation depth as shown in Figure \ref{Fig:Fig5}. %The class \verb|SymbolicStringHandler| used to handle String type constraints is also adapted.
	\item \textbf{Eliminating false alarms}. There exist four kinds of false alarms caused by SSE in SPF: runtime errors in the analyzed program, property violations, user defined exceptions and crashes caused by the program under analysis. We have handled all these issues in our implementation.
\end{itemize}

To use SSE in SPF, users need to configure SPF to use the speculative DFS strategy (using the property  \verb|search.class|) and specify the max speculation depth (using the property \verb|symbolic.speculative.depth|).

\subsection{Experiments}
To evaluate SSE, we have conducted some experiments. The objective of the experiments is to investigate the following research questions.

\textbf{a. Effectiveness and cost}. How about the effectiveness and cost of SSE compared with pure SE? %We measure the effectiveness using two metrics: times of constraint solving and the total running time. We also count the executed instructions and consumed memories to measure the cost.

\textbf{b. Speculation depth}. How does the value of the max speculation depth influence the results and what is the optimal speculation depth for a real-world program?

\textbf{c. Exploration order}. In speculative DFS, the execution tree can be explored from two directions, false-side-first and true-side-first order, which one is better?

\subsubsection{Experimental Setup}
We choose five programs that are often used in the experiments related to JPF \cite{pasareanu2008combining}\cite{visser2006testinput}\cite{staats2010parallel}. \verb|WBS|, the Wheel Brake System, comes from the automotive domain \cite{staats2010parallel}. The rest are all Java data structure programs: red-black tree (\verb|TreeMap|), binary search tree (\verb|BinTree|), binomial heap (\verb|BinHeap|) and Fibonacci heap (\verb|FibHeap|) \cite{visser2006testinput}. In addition, we write a data structure program \verb|List|, which implements a double linked list with sorted elements. %The \verb|List| class supports \verb|add|, \verb|delete|, \verb|find| and \verb|assertOrder| (checking whether the list sorted by keys) operations.
For data structure programs, we use parameterized testing \cite{visser2006testinput}\cite{tillmann2008pex} to generate random call sequences of a limited length. The lines of these programs range from 230 (for \verb|BinTree|) to 477 (for \verb|TreeMap|). The ratios of the infeasible branches in the path spaces range from $0\%$ (for \verb|WBS|) to $42\%$ (for \verb|List|). We choose these programs in our experiments for two reasons. Firstly, these programs are often used in the experiments related to JPF. It is reasonable to choose these programs as the benchmark to evaluate SSE. Secondly, the effectiveness of SSE is heavily influenced by the ratio of infeasible branches in the path space of a program. For our selected programs, the ratios of the infeasible branches cover different levels. In fact, $42\%$ (for \verb|List|) is pretty high. Since each reachable branch has at least one feasible side, this ratio can never be higher than $50\%$ if each branch only has two sides.

We conduct different experiments to investigate the aforementioned research questions. For each program, we perform four kinds of analysis: pure SE with/without optimization and SSE with/without optimization. In each kind of analysis, we vary the value of the max speculation depth and the exploration order independently. For each program, the max speculation depth is increased from 2 to the execution depth of the program. In fact, setting the max speculation depth larger than the execution depth yields the same analysis results as setting that as the execution depth, because in such cases speculation segments always end because of path ending. We use Yices \cite{Yices} as the constraint solver because of its high performance and good usability. All of the experiments are carried out on an Intel Core i7 2.80GHz computer with 8 GB of RAM.

\begin{table*}[t]
%% increase table row spacing, adjust to taste
%\renewcommand{\arraystretch}{1.3}
\renewcommand{\multirowsetup}{\centering}
% if using array.sty, it might be a good idea to tweak the value of
% \extrarowheight as needed to properly center the text within the cells
\caption{Experimental Results (specu. dep.=max speculation depth, B.=Best, W.=Worst)}
\label{Table:Table1}
\centering

%% Some packages, such as MDW tools, offer better commands for making tables
%% than the plain LaTeX2e tabular which is used here.
\begin{tabular}{|c|c c|c|c|c|c|c|c|}\hhline{|=========|}
%\begin{tabular}{|r|r r|r|r|r|r|r|r|}\hhline{|=========|}
\multirow{3}{12mm}{\textbf{Program\\(call seq.\\ length)}} &\multicolumn{2}{c|}{\multirow{3}{12mm}{\textbf{Analysis\\(specu. dep.)}}}  & \multirow{3}{19mm}{\textbf{\#sat/unsat/all\\(Savings)}}
& \multirow{3}{6mm}{\textbf{$\%$\\unsat}} & \multirow{3}{10mm}{\textbf{Search\\Time(s)\\(Savings)}} & \multirow{3}{10mm}{\textbf{Solving\\Time(s)\\(Savings)}} & \multirow{3}{8mm}{\textbf{Solving\\Time\\ratio}} & \multirow{3}{13mm}{\textbf{\#Instruction\\(extra)}} \\
 & \multicolumn{2}{c|}{} & & & & & & \\
 & \multicolumn{2}{c|}{} & & & & & & \\\hhline{|=========|}

 %WBS results
 \multirow{5}{12mm}{\textbf{WBS}} & \multicolumn{2}{c|}{pure SE} & $27646/0/27646$ & 0\% & $66.2$ & $62.9$ & $95\%$ & $1382246$\\
 \cline{2-9}
 & \multirow{2}{6mm}{SSE} & B.($10$) & $14174/0/14174(49\%)$ & $0\%$ & $37.5(43\%)$ & $34.3(45.4\%)$ & $91\%$ & $1382246(0\%)$\\
 & & W.($2$) & $17886/0/17886(35\%)$ & $0\%$ & $44.9(32\%)$ & $41.8(33.5\%)$ & $92\%$ & $1382246(0\%)$\\
 \cline{2-9}
  & SSE+ & B.($10$) & $14174/0/14174(49\%)$ & $0\%$ & $37.3(43.6\%)$ & $34.1(45.7\%)$ & $91\%$ & $1382246(0\%)$\\
  & Opi. & W.($2$) & $17886/0/17886(35\%)$ & $0\%$ & $45(32\%)$ & $42(33.2\%)$ & $93\%$ & $1382246(0\%)$\\\hline

  %Treemap results
  \multirow{5}{12mm}{\textbf{TreeMap\\($5$)}} & \multicolumn{2}{c|}{pure SE} & $27005/17261/44266$ & $39\%$ & $80$ & $74.7$ & $93\%$ & $855119$ \\
  \cline{2-9}
  & \multirow{2}{6mm}{SSE} & B.($2$) & $18569/22045/40614(8\%)$ & $54\%$ & $72.2(9.7\%)$ & $65598(12\%)$ & $91\%$ & $1077553(26\%)$\\
  & & W.($5$) & $20096/23772/43868(1\%)$ & $54\%$ & $79.4(0.8\%)$ & $71515(4\%)$ & $90\%$ & $1548222(81\%)$\\
  \cline{2-9}
  & SSE+ & B.($2$) & $11527/23561/35088(21\%)$ & $67\%$ & $61.1(23.6\%)$ & $54.6(27\%)$ & $89\%$ & $1159619(35.6\%)$\\
  & Opi. & W.($5$) & $13187/23829/37016(16\%)$ & $64\%$ & $66.4(17\%)$ & $59(21\%)$ & $89\%$ & $1549553(81.2\%)$\\\hline

 %Bintree results
  \multirow{5}{12mm}{\textbf{BinTree\\($5$)}} & \multicolumn{2}{c|}{pure SE} & $22381/15589/37970$ & $41\%$ & $76.6$ & $72.2$ & $94\%$ & $381092$\\
  \cline{2-9}
  & \multirow{2}{6mm}{SSE} & B.($2$) & $15913/19215/35128(7.5\%)$ & $55\%$ & $70.5(8.1\%)$ & $65.4(9.4\%)$ & $92\%$ & $578416(52\%)$\\
  & & W.($6$) & $16841/20975/37816(0.4\%)$ & $55\%$ & $77(-0.5\%)$ & $70.8(2\%)$ & $92\%$ & $980918(157.4\%)$\\
  \cline{2-9}
  & SSE+ & B.($2$) & $9191/20086/29277(23\%)$ & $69\%$ & $57.7(25\%)$ & $52.4(27.4\%)$ & $91\%$ & $677685(78\%)$\\
  & Opi. & W.($10$) & $9860/20998/30858(19\%)$ & $68\%$ & $61.6(20\%)$ & $55.7(22.8\%)$ & $90\%$ & $984040(158\%)$\\\hline

 %binheap results
  \multirow{5}{12mm}{\textbf{BinHeap\\($6$)}} & \multicolumn{2}{c|}{pure SE} & $164116/23576/187692$ & $13\%$ & $410$ & $371$ & $90\%$ & $21809086$\\
  \cline{2-9}
  & \multirow{2}{6mm}{SSE} & B.($21$) & $114948/38188/153136(18.4\%)$ & $25\%$ & $335.9(18.1\%)$ & $292.3(21\%)$ & $87\%$ & $29950152(37.3\%)$ \\
  & & W.($2$) & $125178/32932/158110(15.8\%)$ & $21\%$ & $345.6(15.7\%)$ & $306.8(17\%)$ & $89\%$ & $24138598(10.7\%)$\\
  \cline{2-9}
  & SSE+ & B.($21$) & $96600/38202/134802(28.2\%)$ & $28$\% & $300(26.8\%)$ & $257.5(30.6\%)$ & $79\%$ & $29950152(37.3\%)$\\
  & Opi. & W.($2$) & $102410/34164/136574(27.2\%)$ & $25\%$ & $303.9(25.9\%)$ & $264.9(28.6\%)$ & $81\%$ & $25766426(18.1\%)$\\\hline

 %fibheap results:
 \multirow{5}{12mm}{\textbf{FibHeap\\($6$)}} & \multicolumn{2}{c|}{pure SE} & $58014/9142/67156$ & $14\%$ & $148.5$ & $133$ & $90\%$ & $8098034$\\
 \cline{2-9}
 & \multirow{2}{6mm}{SSE} & B.($2$) & $44498/10898/55396(18\%)$ & $20\%$ & $125.2(16\%)$ & $110(17\%)$ & $88\%$ & $8416826(3.9\%)$\\
 & & W.($8$) & $40302/15848/56150(16\%)$ & $28\%$ & $130(13\%)$ & $112.6(15\%)$ & $87\%$ & $10731504(32.5\%)$ \\
 \cline{2-9}
 & SSE+ & B.($2$) & $37694/11906/49600(26\%)$ & $24\%$ & $113(23.9\%)$ & $97.5(26.7\%)$ & $86\%$ & $8859148(9.4\%)$\\
 & Opi. & W.($10$) & $33896/16160/50056(25\%)$ & $32\%$ & $117(21.2\%)$ & $100(24.8\%)$ & $85\%$ & $10731504(32.5\%)$\\
 \hline

 %list resutls
  \multirow{5}{12mm}{\textbf{List\\($6$)}} & \multicolumn{2}{c|}{pure SE} & $128076/94380/222456$ & $42\%$ & $520.6$ & $501.5$ & $96\%$ & $2842969$\\
  \cline{2-9}
  & \multirow{2}{6mm}{SSE} & B.($2$) & $104299/108116/212415(5\%)$ & $51\%$ & $489.3(6\%)$ & $467.7(7\%)$ & $96\%$ & $3832245(34.8\%)$\\
  & & W.($7$) & $118384/121056/239440(-7\%)$ & $51\%$ & $561.4(-8\%)$ & $533.6(-6\%)$ & $95\%$ & $7311109(157.2\%)$\\
  \cline{2-9}
  & SSE+ & B.($2$) & $33488/116635/150123(32.5\%)$ & $78\%$ & $325(37.6\%)$ & $303(39.6\%)$ & $93\%$ & $5371823(89\%)$\\
  & Opi. & W.($20$) & $38705/121176/159881(28.1\%)$ & $76\%$ & $354.1(32\%)$ & $327.7(34.7\%)$ & $93\%$ & $7333909(157.9\%)$\\
  \hhline{|=========|}
\end{tabular}
\end{table*}

\subsubsection{Results}

\noindent
\ \\ \textbf{a. Effectiveness and Cost}

Table \ref{Table:Table1} shows part of the experimental results of three kinds of analysis: pure SE, SSE and SSE with optimization. The first column shows the name of each program associated with its corresponding call sequence length if any. We only list the best case and the worst case of SSE (measured by the search time) when the max speculation depth varies from $2$ to the maximum value. The corresponding max speculation depth is shown after the notation `B.' and `W.'. The third and fourth columns show the numbers of different constraint solving results and the percentage of unsat results respectively. Columns 5 and 6 show the total search time and the percentage of the time spent on constraint solving (the average of three runs). The executed instructions are presented in the last column to show the cost of SSE. All the results shown in Table 1 are collected under the true-side-first exploration order.

SSE (without optimization) performs best for \verb|WBS|, which has no infeasible branches in the execution tree. SSE reduces the times of constraint solving by $49\%$ in the best case and $35\%$ in the worst case. The search time is saved by $43.6\%$ and $32\%$ respectively. SSE performs worst for the program \verb|List|. In the best case, SSE reduces $5\%$ of the times of constraint solving and $6\%$ of the search time. In the worst case, SSE brings extra $7\%$ of the times of constraint solving and $8\%$ of the search time. The reason is that, the high ratio of infeasible branches ($42\%$) causes too many failed speculations, which negate the benefit brought by successful speculations. In average, SSE (without optimization) reduces the search time by $16.8\%$ in the best case and by $8.8\%$ in the worst case.

SSE with optimization outperforms SE and SSE for all programs. The optimization brings the most benefits for \verb|List|, making SSE reduce $32.5\%$ of the times of constraint solving and $37.6\%$ of the search time in the best case. This is because the high ratio of unsat branches provides a lot of chances to perform optimization. As expected, for \verb|WBS|, our optimization brings no benefit because no infeasible branches can be used. In average, SSE with optimization reduces $30\%$ of the times of constraint solving and $30\%$ of the search time in the best case, and $25\%$ of the times of constraint solving and $24.7\%$ of the search time in the worst case.

The results in column 7 shows that, in both of pure SE and SSE, constraint solving dominates most of the search time. The percentage of the time spent on constraint solving in the search time is reduced slightly by SSE.

The last column shows the number of executed instructions in different analysis. We can see that, despite executing a plenty of extra instructions, SSE is still faster than pure SE. Another important point is that SSE nearly does not consume extra memories than pure SE. The reason is that speculative DFS only spends extra memories to store the states in failed speculation segments, which can be ignored in our experiments.

\noindent
\textbf{b. Speculation Depth}

\begin{figure}[t]
\centering
\includegraphics[width=2.4in]{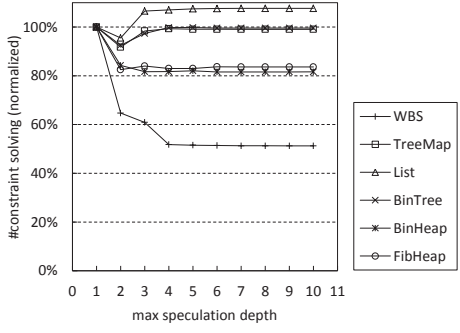}
\caption{Impact of Max Speculation Depth in SSE}
\label{Fig:Fig8}
\end{figure}

\begin{figure}[t]
\centering
\includegraphics[width=2.4in]{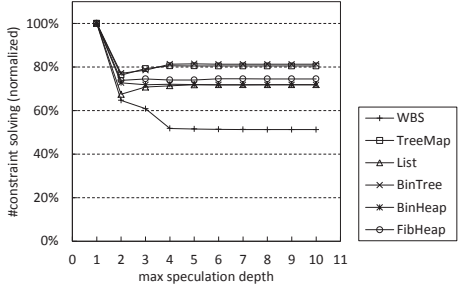}
\caption{Impact of Max Speculation Depth in SSE With Optimization}
\label{Fig:Fig9}
\end{figure}

Figure \ref{Fig:Fig8} shows how the max speculation depth impacts the times of constraint solving in SSE (without optimization). Results for larger speculation depths are omitted since they are nearly the same as the tails of the lines. For the program $P$, let $T_P^p$ be the times of constraint solving in pure SE, and let $T_P^k$ be the times of constraint solving in SSE with the max speculation depth $k$. Figure \ref{Fig:Fig8} shows the result of $T_P^k/T_P^p \times 100\%$.
%The leftmost point is the baseline in the figure, which corresponds to the times of constraint solving in pure SE.
For \verb|List|, \verb|TreeMap|, \verb|BinTree| and \verb|FibHeap|, the optimal speculation depth is 2. Particularly, for \verb|List|, SSE brings benefit only when the max speculation depth is 2. This is because the high ratio of infeasible branches causes too many backtrackings. For \verb|BinHeap|, the optimal speculation depth is 6. For the program without infeasible branches (\verb|WBS|), the results decrease monotonously with the increase of max speculation depth since the speculations never fail.
Figure \ref{Fig:Fig9} shows the impact of the max speculation depth in SSE with optimization, in which the optimal speculation depths for different programs are the same as that in Figure \ref{Fig:Fig8}. Generally, regardless of the tiny fluctuation in the tail, the optimal speculation depth ranges from 2 to 6 and shifts from small to big when the ratio of infeasible branches decreases.
%Since the ratios of the infeasible branches of the programs in our experiments cover different levels, we believe that it is appropriate to set the max speculation depth as 2 for most programs.

We can see that, the optimization technique improves SSE significantly. Another interesting observation is that the results become stable when the max speculation depth reaches a threshold. This also demonstrates that our backtracking mechanism is quite efficient. Besides, the impacts of the max speculation depth on the search time are not shown because they are nearly the same as that on the times of constraint solving.
% the binary search converges quickly in backtracking.

\noindent
\textbf{c. Exploration Order}

\begin{figure}[t]
\centering
\includegraphics[width=2.5in]{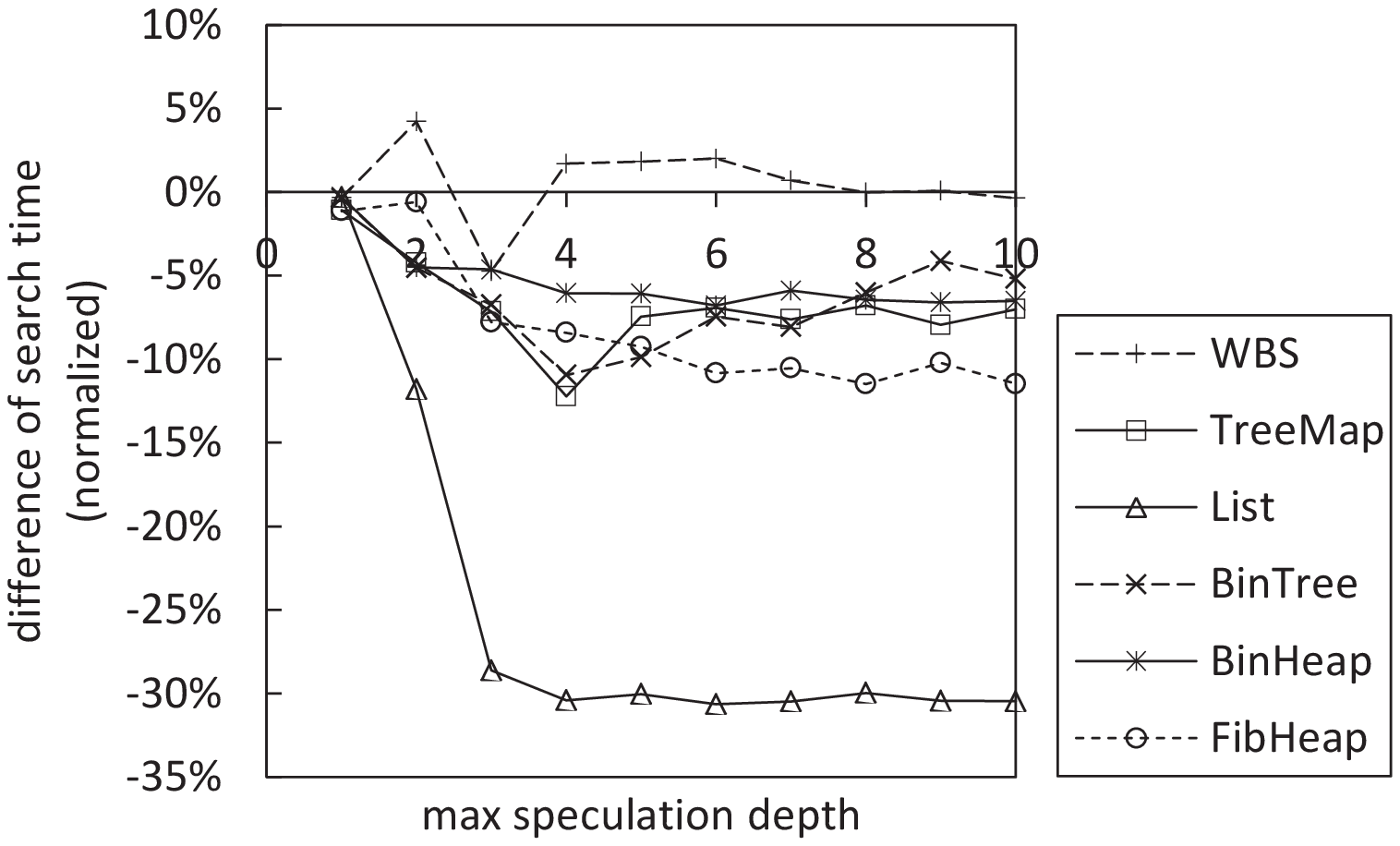}
\caption{Difference of Search Time Between Different Exploration Orders in SSE}
\label{Fig:Fig10}
\end{figure}

\begin{figure}[t]
\centering
\includegraphics[width=2.4in]{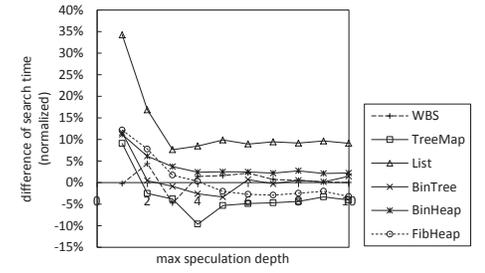}
\caption{Difference of Search Time Between Different Exploration Orders in SSE With Optimization}
\label{Fig:Fig11}
\end{figure}

%As discussed in section 3, our method is affected by the exploration orders.
Figure \ref{Fig:Fig10} illustrates the difference of the search time under two different exploration orders in SSE without optimization. For a program $P$, let $t^p_f$ be the search time of pure SE in false-side-first order, $t^k_t$ be the search time of SSE with true-side-first order and  $t^k_f$ be the search time of SSE with false-side-first order, where $k$ is the max speculation depth. Figure \ref{Fig:Fig10} shows the result of $(t_f^k-t_t^k)/t_s^p \times 100\%$. We can observe that there exists a distinct advantage of false-side-first order. Figure \ref{Fig:Fig11} shows the same calculation under SSE with optimization. In this case, the true-side-first order is slightly better than the false-side-first order, especially when the speculation depth is set as the optimal value.

\begin{table*}[t]
%% increase table row spacing, adjust to taste
%\renewcommand{\arraystretch}{1.3}
% if using array.sty, it might be a good idea to tweak the value of
% \extrarowheight as needed to properly center the text within the cells
\caption{\#Constraint Solving Results of Two Sides}
\label{Table:Table2}
\centering
%% Some packages, such as MDW tools, offer better commands for making tables
%% than the plain LaTeX2e tabular which is used here.
\begin{tabular}{lllllllll}\hline\hline%{|=========|}
\multirow{3}{12mm}{Program} & \multirow{3}{12mm}{\#feasible\\true sides} & \multirow{3}{16mm}{\#infeasible\\true sides(\%)} & \multirow{3}{12mm}{\#feasible\\false sides} & \multirow{3}{16mm}{\#infeasible\\false sides(\%)} &
\multirow{3}{13mm}{\#feasible\\true sides\\of equation} &
\multirow{3}{18mm}{\#infeasible\\true sides\\of equation(\%)\\} &
\multirow{3}{13mm}{\#feasible\\false sides\\of equation} &
\multirow{3}{18mm}{\#infeasible\\false sides\\of equation(\%)}
\\
 & & & & & & & & \\
 & & & & & & & & \\\hline
 WBS     & $13823$    & $0(0\%)$        & $13823$ & $0(0\%)$      &
             $3005$   & $0(0\%)$        & $3005$  & $0(0\%)$     \\%\hline
 List    &	$30276$   & $80952(73\%)$   & $97800$ & $13428(12\%)$ &
            $24660$   & $35328(58.9\%)$ & $46560$ & $13428(22.4\%)$\\%\hline
 TreeMap & $11567$    & $10566(48\%)$   & $15438$ & $6695(30\%)$  &
           $5601$     & $7048(55.7\%)$  & $9484$  & $3165(25\%)$ \\%\hline
 BinTree & $9214$     & $9771(52\%)$    & $13167$ & $5818(31\%)$  &
           $3839$     & $6095(61.4\%)$  & $7793$  & $2141(21.6\%)$ \\%\hline
 BinHeap & $70270$    & $23576(25\%)$   & $93846$ & $0(0\%)$      &
           $0$        & $0(-)$          & $0$     & $0(-)$ \\%\hline
 FibHeap & $25006$    & $8572(26\%)$    & $33008$ & $570(2\%)$    &
           $0$        & $0(-)$          & $0$     & $0(-)$ \\\hline%\hline%{|=========|}
\end{tabular}
\end{table*}

To find the reasons, we collect the constraint solving results of the two sides of the branches under pure SE. The results are shown in Table \ref{Table:Table2}. Column 2 to 5 show the constraint solving results of the two sides in the whole execution tree. Column 6 to 9 show the constraint solving results of the branches with equation constraints. We can see that, the true side has a higher probability to be infeasible in comparison with the false side. The reason is twofold. Firstly, for branches with equation constraints, the true sides have a more than two times higher probability to be infeasible than the false sides. The equation constraint makes the path space along the true side narrower. Secondly, the ratio of the infeasible true sides of the branches with inequation constraints are also higher. We argue that this stems from the characteristics of programs. In programming practice, special cases are usually handled in the \verb|then| branch and other cases are put in the \verb|else| branch. It is reasonable to think that the \verb|then| branch is easier to be infeasible. This finding implies that the execution trees of programs are not bilateral symmetry, but tend to incline to the false sides of the branches.

As a result, the higher rate of feasible false side branches endows SSE in false-side-first order with a higher success rate of speculation, while the lower rate of feasible true side branches provides more information to the optimization. In summary, the dissymmetry of the path space of programs makes SSE with optimization in the true-side-first order the optimal in our experiments, since the shape of the execution tree can be leveraged to the utmost extent.

%An interesting observation is that, for \verb|List|, the optimization makes a reverse between the analysis results in two exploration sides. This is because, the high ratio of infeasible branches increases the success rate of speculation, facilitates SSE in false-side-first order (a higher rate of successful speculation), but benefits much more to the optimization with true-side-first order.

\subsection{Threats to Validity}
The main validity problems need to consider in our experiments are threats to the external validity, which include two aspects: the chosen programs and the implementation platform.

We chose 6 programs in our experiments, 5 of which are often used in the experiments related to JPF. The characteristics of the path spaces of these programs influence the results definitely and our selected programs may not be representative. The ratios of the infeasible branches in our chosen programs range from $0$ to $42\%$. From this perspective, our subjects are quite representative. We limit the call sequences length for data structure programs to control the running time of the experiments. Longer bounds would make constraint solving more time-consuming and may affect the results. The conditions of some branches in data structure programs are heap constraints. We do not perform speculation for heap constraints because the subsequent instructions heavily depends on the condition. We believe that for other types of programs, such as numerical programs or control programs (\verb|WBS| in our experiments), the results may be better.

We selected SPF as the implementation platform and Yices as the constraint solver. A different selection may yield different running time, whereas the times of constraint solving would not change.

\section{Related Work}
Our work is inspired by the speculation execution used in pipelined processors \cite{Smith1981}, which predicts the outcome of a branch and issues the subsequent instructions before the actual branch outcome is known. SSE also executes the instructions after a branch before the feasibility of the branch is known. That is why we use the term \emph{speculative symbolic execution} in this paper.

Speculation is used to improve performance in many other systems, such as operating systems \cite{wester2011operating}, distributed file systems \cite{nightingale2005speculative}. An essential difference between these systems and the work proposed in this paper is that, the performance improvement brought by our method stems from the decrease of the execution times of special operations, rather than the better parallelization brought by speculation in other systems mentioned above. To the best of our knowledge, we are the first to conduct systematic research on using speculation in symbolic execution.

Our work is also related to the large body of work on the scalability problem in symbolic execution, which stems from two reasons: path explosion problem and constraint solving overhead. To attack the path explosion problem, researchers have proposed to use path pruning \cite{bardin2009pruning}\cite{chipounov2009selective}\cite{burnim2008heuristics}, compositional method \cite{godefroid2007compositional}, abstraction \cite{anand2009SEwithabstraction}, state merging \cite{kuznetsov2012statemerge}, parallelism \cite{ciortea2010cloud9}\cite{staats2010parallel} and so on to improve path exploration. To alleviate the constraint solving overhead, a plenty of work have been proposed \cite{cadar2008klee}\cite{cadar2008exe}\cite{sen2005cute}\cite{godefroid2008sage}\cite{puasuareanu2011mixedsolving}\cite{godefroid2005dart}\cite{tillmann2008pex}\cite{erete2011optimizing}. Generally, the optimization techniques employed in current symbolic execution systems attack the constraint solving overhead by query simplification, reusing previous results or fast checking before constraint solving \cite{cadar2008exe}\cite{cadar2008klee}\cite{sen2005cute}\cite{godefroid2008sage}.

%In EXE\cite{cadar2008exe}, besides \emph{Constraint independence}, \emph{Constraint caching} is employed to reuse previous constraint solving results. KLEE\cite{cadar2008klee} uses a number of optimizations to simplify or eliminate queries. CUTE\cite{sen2005cute} uses similar constraint optimizations to improve performance. SAGE\cite{godefroid2008sage} not only employs the common optimizations in dynamic test generation, but also uses \emph{constraint subsumption optimization} to syntactically check the implication relation between constraints before they reach the constraint solver.
%In general, these works attack the constraint solving overhead by query simplification, reusing previous results or fast checking before constraint solving.

The work proposed in this paper is an orthogonal and complementary approach.
SSE employs a new fashion of path exploration technique, aiming to attack the constraint solving overhead by reducing the invocation times of constraint solver. SSE neither reduces the complexity of the queries submitted to the solver, nor caches constraints to reuse previous constraint solving results.
SSE reduces the constraint solving overhead from a unique perspective.
 %by solving the constraints in batches. %Basically, SSE can be used in any classical symbolic executors with the optimization techniques mentioned above together.

Lei Bu \emph{et al.} use the idea of speculation in \cite{LeiBu2011} for the reachability checking of linear hybird automata. Different from their work, the speculation in SSE is limited by a specific number, but the speculation in \cite{LeiBu2011} stops when a target location is reached, which is more like a target driven `slicing'. This is caused by different contexts of using speculation. Another difference is the backtracking mechanism. We use binary search to find backtracking points, whereas the irreducible infeasible set technique \cite{chinneck1991locating} is employed in \cite{LeiBu2011} for backtracking. For SSE, binary search is stable and effective. Nevertheless, the minimal unsatisfiable core extraction technique \cite{cimatti2011computing} may also be used in the backtracking of SSE to reduce the times of constraint solving further.

%In paper\cite{LeiBu2011}, Lei Bu \emph{et al.} proposed to use Lazy-DFS for reachability checking problem of linear hybrid automata. Lazy-DFS invokes linear programming solver only when the last location of the current visiting path is the target location. They also employ irreducible infeasible set technique\cite{chinneck1991locating} to find backtracking point. Our independently and concurrently developed work has the following two differences. First, in SSE, the constraint solver is invoked when the number of unchecked branches reaches a specific number. While Lazy-DFS is more like the target driven `slicing'. The second difference is that we use binary search in finding backtracking point. The reason is twofold. First, the minimal unsatisfiable core (MUC) extraction in SMT is still in the research phase, despite it is supported by a few SMT solvers\cite{cimatti2011computing}. Second, when a path condition owns more than one MUC, extra computation is needed to locate the backtracking point. Nevertheless, we are going to investigate to use similar techniques to help our backtracking.

At the time of this writing, EPFL released S$^2$E v1.2 \cite{S2E}. An optimization named \emph{speculative forking} is used in the concolic execution, where symbolic states are forked without regard to the feasibility at the branch that depends on symbolic values. These speculatively generated states are used as backtracking points (if feasible) to avoid re-execution from scratch when new inputs are generated. Although we both use the similar term, speculation is used to achieve different goals.
%At the time of this writing, EPFL released S2E v1.2\cite{S2E}, in which \emph{speculative forking} is used to reduce the invocation times of constraint solver in concolic execution. The search procedure forks symbolic states regardless of feasibility and checks the feasibility in the path end.
%Up to now, no publication on this optimization technique has been published.

The philosophy of SSE is a little similar to that used in many static analysis systems, which consider extra program behaviors and eliminate false alarms in the end \cite{zitser2004testing}\cite{jung2005taming}. The difference is that SSE prunes the infeasible behaviors in an appropriate chance to keep results precise and make a good tradeoff between the cost and the benefit as well.

\section{Conclusion and Future Work}
We have proposed a new fashion of symbolic execution named \emph{speculative symbolic execution} to reduce the invocation times of constraint solver, and hence extend the scalability of symbolic execution. SSE attacks the constraint solving overhead, which is almost always the most dominant in the running time of symbolic execution. We have proposed the speculative DFS algorithm and discussed its effectiveness. We also propose a key optimization technique, named \emph{absurdity based optimization}, to further improve SSE. This optimization is very effective especially for the programs with a high ratio of infeasible branches.

We have implemented SSE and our optimization technique on top of SPF. Experiments have been conducted to investigate several important research questions. The experimental results on six programs show that, SSE can reduce the invocation times of constraint solver by $21\%$ to $49\%$ (with a medium of $30\%$), and save the search time from $23.6\%$ to $43.6\%$ (with a medium of $30\%$).
For future work, we plan to research on different search styles and use existing query optimization techniques to enhance SSE further.

\section*{Acknowledgment}

We would like to thank Corina P\v{a}s\v{a}reanu and Willem Visser for their helps and discussions on using JPF and SPF.



\bibliographystyle{IEEEtran}

\bibliography{SSE}
%\begin{thebibliography}{1}


%\end{thebibliography}

\newpage

\section{Appendix A}
\subsection{Proof of Proposition 2}
 We define the height of an execution tree as the number of branches in the longest path in the tree. Let $Tr_n$ be an arbitrary execution tree with height $n$. Let $T^p(Tr_n)$ and $T^k(Tr_n)$ be the times of constraint solving in performing pure SE and speculative DFS (SP-DFS) on $Tr_n$ respectively, where $k$ is the max speculation depth. Proposition 2 claims that for any $Tr_n$, $T^k(Tr_n)>1/2 \times T^p(Tr_n)$. We use induction on the height of the execution tree to prove this proposition.

\begin{figure}[b]
\centering
\includegraphics[width=1in]{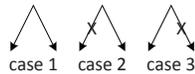}
\caption{Execution Tree With Height 1}
\label{Fig:Fig12}
\end{figure}

 \textbf{Basis}: Fon $n=1$, as shown in Figure \ref{Fig:Fig12}, there are three possible shapes for $Tr_1$. In each case, both of SP-DFS and pure SE need two times of constraint solving. So Proposition 2 holds for $n=1$.

\textbf{Induction step}: Suppose that Proposition 2 holds for $Tr_n$, \emph{i.e.}, $T^k(Tr_n)>1/2 \times T^p(Tr_n)$, we now show $T^k(Tr_{n+1})>1/2 \times T^p(Tr_{n+1})$.

\begin{figure}[b]
\centering
\includegraphics[width=2in]{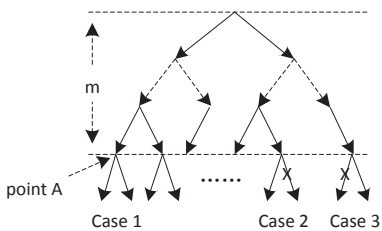}
\caption{Execution Tree With Height $n+1$}
\label{Fig:Fig13}
\end{figure}

As shown in Figure \ref{Fig:Fig13}, $Tr_{n+1}$ can be regarded as constructed by adding a level of branches to some leaves (at least one) of $Tr_n$. Let $e$ be a leaf of $Tr_n$, as shown in Figure \ref{Fig:Fig13}, $e$ can be extended in three different cases. Let $b_l$ and $b_r$ (at least one is feasible) be the two new branches under $e$. In pure SE, each of $b_l$ and $b_r$ need one time of constraint solving, no matter the branch is feasible or not. Now we analyze what difference these new branches bring to the times of constraint solving between exploration of $Tr_{n}$ and $Tr_{n+1}$ by SP-DFS.

\textbf{Case 1: \emph{$b_l$ and $b_r$ are both feasible}}. In SP-DFS, if \emph{point A} (In Figure \ref{Fig:Fig13}) reaches the max speculation depth when exploring $Tr_n$, then in exploration on $Tr_{n+1}$, a new speculation segment starts at $b_l$, so $b_l$ consumes one time of constraint solving.  If \emph{point A} does not reach the max speculation depth, then $b_l$ will be included in the same speculation segment with the branch above \emph{point A}. Since $b_l$ is satisfiable, it brings no extra constraint solving. What's more, to complete $b_r$, SP-DFS backtracks to \emph{point A} and spends one time of constraint solving. In summary, the two new branches brings 1 or 2 extra times of constraint solving in SP-DFS.

\textbf{Case 2: \emph{$b_l$ is feasible and $b_r$ is infeasible}}. The argument is similar to case 1, except that $b_r$ is unsatisfiable.

\textbf{Case 3: \emph{$b_l$ is infeasible and $b_r$ is feasible}}. If \emph{point A} reaches the max speculation depth when exploring $Tr_n$, then when exploring $Tr_{n+1}$, one time of constraint solving is needed for $b_l$. Otherwise, when exploring $Tr_{n+1}$, $b_l$ will be included in the same speculation segment with the branch above \emph{point A}. This speculation fails because $b_l$ is infeasible, and needs $\lceil \log_2k \rceil$ or $\lceil \log_2k \rceil+1$ times of constraint solving in backtracking. For $b_r$, one time of solving is needed. In summary, $b_l$ and $b_r$ bring $\lceil \log_2k \rceil + 1$ or $\lceil \log_2k \rceil+2$ times of constraint solving.

In summary, for each of the three cases above, $b_r$ always needs one time of constraint solving, and $b_l$ brings 0 or more times. Since $b_l$ and $b_r$ need 2 times of constraint solving in pure SE, therefore the increased times of  constraint solving in SP-DFS is larger than half of that in pure SE. According to the induction hypothesis, $T^k(Tr_n) > 1/2 \times T^p(Tr_n)$, we can get $T^k(Tr_{n+1}) > 1/2 \times T^p(Tr_{n+1})$.

We claim that proposition 2 holds for an arbitray execution tree.

\subsection{Proof of Equation (1)}
Let $Tree_n$  be a full binary execution tree of height $n$.
Let $T^k(Tree_n)$ be the times of constraint solving in SP-DFS, where $k$ is the max speculation depth.
There are the following two cases:

\textbf{Case 1:} $n\leq k$.

When $n\leq k$, speculation segments always terminate because of path ending. As a result, constraint solving always occurs at the end of a path, which makes the invocation times of constraint solver equal to the number of paths. So we get

\begin{displaymath}\tag{A.2}
T^k(Tree_n)=2^n (n\leq k)
\end{displaymath}

\textbf{Case 2:} $n>k$.

Now we calculate the relation between $T^k(Tree_n)$ and $T^k(Tree_{n+1})$.

\begin{figure}[t]
\centering
\includegraphics[width=1.8in]{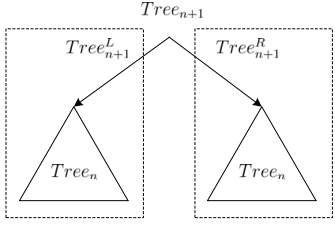}
\caption{A Full Binary Tree With Height of $n+1$}
\label{Fig:Fig14}
\end{figure}

As shown in Figure \ref{Fig:Fig14}, $Tree_{n+1}$ is composed of two symmetrical subtrees, say $Tree^L_{n+1}$  and $Tree^R_{n+1}$ respectively, each of which is constructed by adding a branch on top of a $Tree_n$.

According to the SP-DFS procedure, it is easy to know $T^k(Tree^L_{n+1})=T^k(Tree^R_{n+1})$ and $T^k(Tree_{n+1}) = 2 \times T^k(Tree^L_{n+1})$. Therefore, to calculate the relation between $T^k(Tree_n)$ and $T^k(Tree_{n+1})$, it suffices to know the relation between $T^k(Tree_n)$ and $T^k(Tree_{n+1}^L)$.

\begin{figure}[t]
\centering
\includegraphics[width=1.6in]{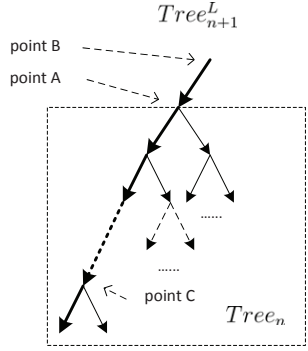}
\caption{Impact of Adding a Branch on Top of $Tree_n$}
\label{Fig:Fig15}
\end{figure}

As shown in Figure \ref{Fig:Fig15}, $Tree^L_{n+1}$ consists of a $Tree_n$ and a branch on the top. Now we focus on what difference this new branch brings to the times of constraint solving in exploring $Tree^L_{n+1}$ and $Tree_n$ by SP-DFS.

For $Tree_n$, SP-DFS starts at \emph{point A} (in Figure \ref{Fig:Fig15}). The first completed path in $Tree_n$ is the leftmost one, say $P_l$, which is marked with thick arrows in Figure \ref{Fig:Fig15}. It is easy to know that when exploring $P_l$, a total of $\lceil n/k \rceil $ times of constraint solving are needed. Besides, the lengths of these speculation segments are all $k$ except the last one, which includes $n\%k$ branches. After complete $P_l$, the search procedure backtracks to \emph{point C} and starts to traverse the other part of $Tree_n$.

For $Tree_{n+1}^L $, SP-DFS starts at \emph{point B}. The leftmost path is still the first to complete, needing a total of $\lceil (n+1)/k \rceil$ times of constraint solving. After completing the leftmost path, SP-DFS backtracks to the \emph{point C} and continues. It is clear that after \emph{point C}, SP-DFS performs identically as that in exploring $Tree_n$, because the backtracking points separate the the leftmost path and the other parts of the tree. Therefore, the only difference in exploring $Tree^L_{n+1}$ happens on the leftmost path.

Since
\begin{displaymath}\tag{A.3}
\lceil(n+1)/k\rceil=
\begin{cases}
\lceil n/k\rceil &  (n \% k \neq 0)\\
\lceil n/k\rceil + 1 & (n \% k = 0)
\end{cases}
\end{displaymath}
The relation between  $T^k(Tree_{n+1})$ and $T^k(Tree_n)$ can be quantified by the following equation:
\begin{displaymath}\tag{A.4}
T^k(Tree^L_{n+1})=
\begin{cases}
T^k(Tree_n) & (n\%k \neq 0)\\
T^k(Tree_n)+1 & (n\%k = 0)\\
\end{cases}
\end{displaymath}

From Equation A.4, it is not hard to get the following equation, which quantifies the times of constraint solving in performing SP-DFS on $Tree_n$.
\begin{displaymath}\tag{A.5}
T^k(Tree_n)=
\begin{cases}
2^n &  (n \leq k)\\
2^n + \dfrac{2^n-2^{(n\%k)}}{2^k-1} & (n > k)
\end{cases}
\end{displaymath}

\end{document}